\documentclass[preprint,aps,floatfix]{revtex4}
\usepackage{hyperref}
\everymath={\displaystyle}
\usepackage{young}
\usepackage{rotating}
\usepackage{longtable}
\def\1{\mbox{l\hspace{-0.53em}1}}

\usepackage{multirow}
\newlength{\AccoHaut}

\begin{document}
\title{Negative parity baryons in the $1/N_c$ expansion: the  quark excitation
versus the meson-nucleon resonance picture }

\author{N. Matagne\footnote{E-mail address: nicolas.matagne@umons.ac.be}}
\affiliation{University of Mons, Service de Physique Nucl\'eaire et 
Subnucl\'eaire,Place du Parc 20, B-7000 Mons, Belgium}

\author{Fl. Stancu\footnote{E-mail address: fstancu@ulg.ac.be}}
\affiliation{University of Li\`ege, Institute of Physics B5, Sart Tilman,
B-4000 Li\`ege 1, Belgium}

\date{\today}

\begin{abstract}
In order to better understand the fundamental issue regarding the compatibility between the quark-shell picture 
and that of resonances in meson-nucleon scattering in large $N_c$ QCD
we extend the work of Cohen and Lebed on mixed symmetric $\ell $ = 1 baryons to analyze excited states 
with $\ell $ = 3.  We give an explicit proof on the degeneracy of mass eigenvalues  
of a simple Hamiltonian including operators up to order $\mathcal{O}(N^0_c)$ \emph{i.e.} 
neglecting $1/N_c$ corrections in the quark-shell picture in the large $N_c$ limit.  
We obtain three sets of degenerate states with $\ell = 3$,
as in the case of   $\ell $ = 1 baryons.  The compatibility between this picture 
and that of  resonances in meson-nucleon scattering is discussed in the light of
the present results.
\end{abstract}

\maketitle

 

\section{Introduction}
The usefulness of the $1/N_c$ expansion method \cite{HOOFT,WITTEN,Gervais:1983wq,DM} to describe ground state baryons 
has been clearly demonstrated  \cite{Jenk1,DJM94,DJM95}. It is based on the contracted spin-flavor symmetry SU(2$N_f$) 
that emerges in large $N_c$ QCD.   
For excited baryons the problem is more involved and needs more investigations.

There are two complementary pictures of large $N_c$ for the baryon resonances. Using the terminology of Refs. \cite{Pirjol:2003ye,COLEB1,COLEB2} these are:

(i) the  SU(2$N_f$) $\times$ O(3) called {\it the quark-shell picture}  where the role of O(3) is essentially to include orbital excitations. 
This picture allows to classify baryons in excitation bands $N$, like in the quark model \cite{Matagne:2005gd,Semay:2007cv}. 
For $N_c$ = 3 each band contains a number of SU(6) $\times$ O(3) multiplets.  
A practical and commonly used procedure is to consider an excited state as a single quark excitation about a spin-flavor
symmetric core \cite{Goi97,CCGL}. The $N_c$ counting is implemented by introducing operators that break the  SU(2$N_f$)
symmetry in powers of $1/N_c$.  The coefficients of these operators encode the quark dynamics and
are fitted from experiment. 

(ii)   the {\it resonance or scattering picture} 
derived from symmetry features shared by various chiral soliton 
models. 
The role of large $N_c$ QCD is to relate the scattering amplitudes in various channels with $K$-amplitudes, where $K$
is the grand spin $\vec{K} = \vec{I} + \vec{J}$. 
These are  linear relations in the meson-nucleon scattering amplitudes from which one can infer some patterns of 
degeneracy among resonances. 

For the nonstrange lowest negative parity baryons belonging to the $[70,1^-]$ multiplet 
 (the $N = 1$ band in quark model terms )  Cohen  and Lebed  have shown \cite{COLEB1}
 that the two pictures share the same pattern 
of degeneracy from which they concluded that the two pictures are generically compatible.
Simultaneously with Cohen and Lebed,  Pirjol and Schat \cite{Pirjol:2003ye} found 
the same sets of degenerate states, corresponding to irreducible representations of the contracted SU(4)$_c$
symmetry and  the three degenerate multiplets obtained by them were called three towers of states.
Moreover, to the three leading-order operators in the mass formula they added $1/N_c$
corrections and reanalyzed the mass spectrum of the lowest negative parity nonstrange 
baryons.  They found  ambiguities in the identification of physical states with $N_c$ = 3 
with the degenerate large $N_c$ tower states.

One should also mention that in the SU(4) case, 
prior to Refs. \cite{COLEB1,Pirjol:2003ye},  the degeneracy of multiplets corresponding 
to irreducible representations of the contracted SU(4)$_c$ symmetry
was first discussed  by Pirjol and Yan in Ref. \cite{PY2}.

Later on, the compatibility between the two pictures was discussed  on a general basis   
again by Cohen and Lebed  \cite{COLEB2}. By full compatibility it was understood   that any complete spin-flavor multiplet 
within one picture fills  the quantum numbers in the other picture.
The analysis involved group theoretical arguments and  the nature 
of quark excitations in a hedgehog picture.  
The compatibility was generally claimed for completely symmetric, mixed symmetric  and completely
antisymmetric  states of $N_c$ quarks having angular momentum up to $\ell = 3$.  However an explicit proof 
regarding the degeneracy of mass eigenvalues in the quark-shell picture is known only for $\ell = 1$
\cite{Pirjol:2003ye,COLEB1}. For symmetric states with $\ell = 0$ and  2  it is inferred from  previous studies   \cite{Jenk1,DJM94,DJM95}
and \cite{Goity:2003ab} respectively. Similar arguments hold for symmetric $\ell$ = 4 states as well
\cite{Matagne:2004pm}.

The aim of the present work is to give an explicit analytical proof of the degeneracy 
of mass eigenvalues in the quark-shell picture for $\ell$ = 3 and present  its pattern 
of degeneracy as compared to that of the meson-nucleon scattering picture.  For this purpose we use the
Hamiltonian of Ref. \cite{COLEB1} and calculate all the possible eigenvalues.  We find a pattern of 
degeneracy which is compared to that given in Ref. \cite{COLEB2} from general arguments and
discuss the compatibility of the two pictures explicitly.

\section{ The quark-shell  picture }
In the quark-shell  picture the authors of Ref. \cite{COLEB1} start from  the leading-order Hamiltonian 
including operators up to order $\mathcal{O}(N^0_c)$ which has the following form \cite{CCGL}
\begin{equation}\label{TOY}
H = c_1 \ \1  + c_2 \ell \cdot s + c_3 \frac{1}{N_c}\ell^{(2)} \cdot g \cdot G_c
\end{equation}
This operator is defined in the spirit of a Hartree picture (mean field) \cite{WITTEN}
where the matrix elements of the first term are $N_c$ on all baryons, and the spin-orbit term $\ell \cdot s$ 
which is a one-body operator and the  third term - a two-body operator containing the tensor  
$\ell^{(2)ij}$ of O(3) - have matrix elements of order $\mathcal{O}(N^0_c)$. The neglect of $1/N_c$
corrections  in the $1/N_c$ expansion makes sense for the comparison with the scattering picture
in the large $N_c$ limit, as described in the following section.

We remind that the SU(4) generators are $S^i$, $T^a$ and $G^{ia}$
and $\ell^i$ are the O(3) generators which form the tensor operator $\ell^{(2)ij} = 1/2 ~\{\ell^i , \ell^j \} - 1/3~ \ell^2 ~\delta_{i,-j}$.
In the manner of Ref. \cite{CCGL} they are
decomposed into two parts, one acting on the excited quark and the other on the
ground state core.  Thus $\ell^i$, $s^i$, $t^a$ and $g^{ia}$ act on the excited quark
and $S^i_c$, $T^a_c$ and $G^{ia}_c$ act on the ground state core. Starting from the Hamiltonian (\ref {TOY}) 
the authors of Ref. \cite{COLEB1} show that the masses of the $[70, 1^-]$ multiplet 
described by this model to leading order in the $1/N_c$ expansion
fall into three degenerate multiplets given by three distinct masses denoted by $m_0$, $m_1$ and $m_2$,
which are linear in the parameters $c_1$, $c_2$ and  $c_3$ . Their expression are
\begin{equation}\label{clmass}
m_0 = c_1 N_c - (c_2 + \frac{5}{24} c_3), \; \; \; m_1 =  c_1 N_c - \frac{1}{2}(c_2 - \frac{5}{24} c_3), 
\; \; \; m_2 =  c_1 N_c + \frac{1}{2}(c_2 - \frac{1}{24} c_3).
\end{equation}

From its form one can see that the  Hamiltonian  (\ref{TOY}) incorporates the property 
that the characteristic $N_c$ scaling for
the excitation energy of baryons is $N^0_c$   \cite{WITTEN}.

The spectrum obtained from the Hamiltonian (\ref{TOY}) is identical to that derived by Pirjol and Schat \cite{Pirjol:2003ye}
for $\ell$ = 1. Note that the third operator of Ref. \cite{Pirjol:2003ye} contains an extra factor of 3, which should be 
taken into account when comparing the eigenvalues.

Below we give the mass matrices obtained from the Hamiltonian (\ref{TOY}) for $\ell$ = 3.  As we shall see,
this Hamiltonian has the remarkable property that for $\ell$ = 3 as well, its eigenvalues are simple 
linear expressions in  the coefficients  $c_i$,  which makes the discussion very convenient.

\subsection{The nucleon case}
We have the following $[N_c-1,1]$ spin-flavor  ($SF$) states which form a symmetric state with the orbital $\ell$ = 3 state of partition $[N_c - 1,1]$ as well  

\begin{enumerate}
\item
$\left[N_c - 1, 1\right]_{SF} = \left[\frac{N_c+1}{2}, \frac{N_c - 1}{2}\right]_{S} \times  \left[\frac{N_c+1}{2}, \frac{N_c - 1}{2}\right]_{F} $, $N_c  \geq 3$ \\
with $S = 1/2$ and $J = 5/2, 7/2$
\item
$\left[N_c - 1, 1\right]_{SF} = \left[\frac{N_c+3}{2}, \frac{N_c - 3}{2}\right]_{S} \times  \left[\frac{N_c+1}{2}, \frac{N_c - 1}{2}\right]_{F} $,   $N_c  \geq 3$ \\  
with $S = 3/2$ and $J = 3/2, 5/2, 7/2, 9/2$.
\end{enumerate} 
They give rise to matrices of a given $J$ either $2 \times 2$ or $1 \times 1$. States of symmetry $[N_c - 1, 1]_{SF}$ with  
  $S = 5/2$,  like for $\Delta$ (see below), which together with $\ell = 3$ could give rise  to $J = 11/2$, 
   are not allowed for $N$, by inner products of the permutation group
  \cite{Stancu:1991rc}. Therefore the resonance $N_{11/2}$ should belong to the $N = 5$ band ($\ell$ = 5), as suggested
in the before last section. 
For $N_c$ = 3 the above states belong to the $^28$ and $^48$ multiplets of SU(2) $\times$ SU(3) respectively.
  
We calculate the matrix elements using the formulas from Appendix A. The expectation value for  the  $\ell$ = 3  
$N_{3/2}$  state is 
\begin{equation}  
m^{(1)}_{N_{3/2}} = c_1 N_c - 2 c_2 - \frac{3}{4} c_3 
\end{equation}

The matrix for $N_{5/2}$ is 
 \begin{eqnarray}
\label{N5halves}
M^{\ell=3}_{N_{5/2}} & = &
 \left( 
 \begin{array}{cc}
c_1 N_c -\frac{4}{3} c_2  &~~~~~~ - \frac{\sqrt{5}}{3}  c_2 -  \frac{3 \sqrt{5}}{8} c_3 \\
- \frac{\sqrt{5}}{3}  c_2 -   \frac{3 \sqrt{5}}{8} c_3  &~~~~~~   c_1 N_c - \frac{7}{6} c_2 + \frac{3}{16} c_3  
\end{array} 
\right) 
\end{eqnarray}
Its eigenvalues are 
\begin{equation}  
m^{(1)}_{N_{5/2}} = c_1 N_c - 2  c_2 - \frac{3}{4} c_3 
\end{equation}
\begin{equation}  
m^{(2)}_{N_{5/2}} = c_1 N_c - \frac{1}{2} c_2 +\frac{15}{16} c_3 
\end{equation}

The matrix for $N_{7/2}$ is 
\begin{eqnarray}
\label{N7halves}
M^{\ell=3}_{N_{7/2}} & = &
 \left( 
 \begin{array}{cc}
c_1 N_c + c_2  &~~~~~~ - \frac{\sqrt{3}}{2}  c_2 +  \frac{5 \sqrt{3}}{16} c_3 \\
 - \frac{\sqrt{3}}{2}  c_2 +  \frac{5 \sqrt{3}}{16} c_3 &~~~~~~   c_1 N_c +  \frac{5}{8} c_3  
\end{array} 
\right)
\end{eqnarray}
and its  eigenvalues are
\begin{equation}  
m^{(1)}_{N_{7/2}} = c_1 N_c - \frac{1}{2} c_2 + \frac{15}{16} c_3 
\end{equation}
\begin{equation}  
m^{(2)}_{N_{7/2}} = c_1 N_c + \frac{3}{2} c_2 - \frac{5}{16} c_3 
\end{equation}

The expectation value of the $N_{9/2}$ is 
\begin{equation}  
m^{(1)}_{N_{9/2}} = c_1 N_c + \frac{3}{2} c_2 - \frac{5}{16} c_3 
\end{equation}

\subsection{The $\Delta$ case}
We have the following basis states in the spin-flavor space compatible with the orbital state $[N_c -1,1]$ with $\ell$ = 3

\begin{enumerate}
\item
$\left[N_c - 1, 1\right]_{SF} = \left[\frac{N_c+1}{2}, \frac{N_c - 1}{2}\right]_{S} \times  \left[\frac{N_c+3}{2}, \frac{N_c - 3}{2}\right]_{F} $,   $N_c  \geq 3$ \\  
with $S = 1/2$ and $J = 5/2, 7/2$,
\item
$\left[N_c - 1, 1\right]_{SF} = \left[\frac{N_c+3}{2}, \frac{N_c - 3}{2}\right]_{S} \times \left[\frac{N_c+3}{2}, \frac{N_c - 3}{2}\right]_{F} $,   $N_c  \geq 5$ \\  
with $S = 3/2$ and $J = 3/2, 5/2, 7/2, 9/2$,
\item
$\left[N_c - 1, 1\right]_{SF} = \left[\frac{N_c+5}{2}, \frac{N_c - 5}{2}\right]_{S} \times \left[\frac{N_c+3}{2}, \frac{N_c - 3}{2}\right]_{F} $,   $N_c  \geq 7$ \\ 
with $S = 5/2$ and $J = 1/2, 3/2, 5/2, 7/2, 9/2, 11/2$.
\end{enumerate} 
As above, they indicate the size of a matrix of fixed $J$.  For $N_c = 3$ the first state belongs to the $^210$ multiplet.
The other two types of states do not appear in the real world with $N_c = 3$.
Note that both for $N_J$ and $\Delta_J$ states the size of a given matrix equals the multiplicity of the corresponding
state indicated in Table 1 of Ref. \cite{COLEB2} for $\ell = 3$. 

The expectation value for  the   
$\Delta_{1/2}$  state is 
\begin{equation}  
m^{(1)}_{\Delta_{1/2}} = c_1 N_c - 2 c_2 - \frac{3}{4} c_3 
\end{equation}  
The matrix for $\Delta_{3/2}$ is 
 
\begin{eqnarray}
\label{delta3halves}
M^{\ell=3}_{\Delta_{3/2}} & = &
 \left( 
 \begin{array}{cc}
c_1 N_c - \frac{4}{5} c_2 +  \frac{3}{5} c_3 &~~~~~~ - \frac{3}{5}  c_2 -  \frac{27}{40} c_3 \\
- \frac{3}{5}  c_2 -  \frac{27}{40} c_3 &~~~~~~   c_1 N_c - \frac{17}{10} c_2 -  \frac{33}{80} c_3  
\end{array} 
\right)
\end{eqnarray}
The eigenvalues of this matrix are
\begin{equation} 
m^{(1)}_{\Delta_{3/2}} = c_1 N_c - 2 c_2 - \frac{3}{4} c_3
\end{equation}
\begin{equation}  
m^{(2)}_{\Delta_{3/2}} = c_1 N_c - \frac{1}{2} c_2 + \frac{15}{16} c_3 
\end{equation}

For $\Delta_{5/2}$ we have 
\begin{eqnarray}
\label{delta5halves}
M^{\ell=3}_{\Delta_{5/2}} & = &
 \left( 
 \begin{array}{ccc}
c_1 N_c + \frac{2}{3} c_2  &~~~~~~   \sqrt{\frac{1}{2}} ~ ( \frac{5}{3} c_2 -  \frac{3}{8}  c_3 ) 
& ~~~~\frac{3}{4} \sqrt{\frac{1}{2}} c_3 \\
  \sqrt{\frac{15}{2}} ( \frac{1}{2} c_2 + \frac{1}{16}  c_3 ) &~  c_1 N_c - \frac{7}{15} c_2 - \frac{3}{20} c_3 
  &   - \frac{3}{2 \sqrt{2}} c_2 -   \frac{3}{16 \sqrt{2}} c_3 \\ 
\frac{3}{4} \sqrt{\frac{1}{2}} c_3     &  ~~~~ - \frac{3}{2 \sqrt{2}} c_2 -   \frac{3}{16 \sqrt{2}} c_3   
 &  ~~~~~~~  c_1 N_c - \frac{6}{5} c_2  + \frac{1}{40} c_3
\end{array} 
\right)
\end{eqnarray}
The eigenvalues of this matrix are
\begin{equation} 
m^{(1)}_{\Delta_{5/2}} = c_1 N_c - 2 c_2 - \frac{3}{4} c_3
\end{equation}
\begin{equation}  
m^{(2)}_{\Delta_{5/2}} = c_1 N_c - \frac{1}{2} c_2 + \frac{15}{16} c_3 
\end{equation}
\begin{equation}  
m^{(3)}_{\Delta_{5/2}} = c_1 N_c + \frac{3}{2} c_2 - \frac{5}{16} c_3 
\end{equation}

For $\Delta_{7/2}$ we obtain  
\begin{eqnarray}
\label{delta7halves}
M^{\ell=3}_{\Delta_{7/2}} & = &
 \left( 
 \begin{array}{ccc}
c_1 N_c - \frac{1}{2} c_2  &~~~~~~    \sqrt{\frac{15}{2}} ( \frac{1}{2} c_2 +  \frac{1}{16}  c_3 )
& ~~~~\frac{3 \sqrt{15}}{16} c_3 \\
  \sqrt{\frac{15}{2}} ( \frac{1}{2} c_2 + \frac{1}{16}  c_3 ) &~  c_1 N_c - \frac{1}{2} c_3 
  &   - \frac{3}{2 \sqrt{2}} c_2 -   \frac{3}{16 \sqrt{2}} c_3 \\ 
\frac{3 \sqrt{15}}{16} c_3    &  ~~~~ - \frac{3}{2 \sqrt{2}} c_2 -   \frac{3}{16 \sqrt{2}} c_3   
 &  ~~~~~~~  c_1 N_c - \frac{1}{2} c_2  + \frac{3}{8} c_3
\end{array} 
\right)
\end{eqnarray}
and the eigenvalues of this matrix are
\begin{equation} 
m^{(1)}_{\Delta_{7/2}} = c_1 N_c - 2 c_2 - \frac{3}{4} c_3
\end{equation}
\begin{equation}  
m^{(2)}_{\Delta_{7/2}} = c_1 N_c - \frac{1}{2} c_2 + \frac{15}{16} c_3 
\end{equation}
\begin{equation}  
m^{(3)}_{\Delta_{7/2}} = c_1 N_c + \frac{3}{2} c_2 - \frac{5}{16} c_3 
\end{equation}

For $\Delta_{9/2}$ we obtain  
\begin{eqnarray}
\label{delta9halves}
M^{\ell=3}_{\Delta_{9/2}} & = &
 \left( 
 \begin{array}{cc}
c_1 N_c + \frac{3}{5} c_2 +  \frac{1}{4} c_3 &~~~~~~ - \frac{3 \sqrt{11}}{10}  c_2 +  \frac{3 \sqrt{11}}{16} c_3 \\
~~  - \frac{3 \sqrt{11}}{10}  c_2 +  \frac{3 \sqrt{11}}{16} c_3 &~~~~~~~   c_1 N_c - \frac{1}{2} c_2 -  \frac{15}{16} c_3  
\end{array} 
\right)
\end{eqnarray}
with eigenvalues  
\begin{equation}  
m^{(1)}_{\Delta_{9/2}} = c_1 N_c - \frac{1}{2} c_2 + \frac{15}{16} c_3 
\end{equation}
\begin{equation}  
m^{(2)}_{\Delta_{9/2}} = c_1 N_c + \frac{3}{2} c_2 - \frac{5}{16} c_3 
\end{equation}

Finally, the expectation value of the $\Delta_{11/2}$ one component state is
\begin{equation}  
m^{(1)}_{\Delta_{11/2}} = c_1 N_c + \frac{3}{2} c_2 - \frac{5}{16} c_3 
\end{equation}

It is remarkable that the 18 available eigenstates with $\ell$ = 3 fall into three degenerate multiplets,
like for $\ell$ = 1.  If the degenerate masses are denoted by $m'_2$,  $m_3$ and $m_4$ 
we have
\begin{equation}\label{mass3}
m'_2 =   m^{(1)}_{\Delta_{1/2}}  =  m^{(1)}_{N_{3/2}} = m^{(1)}_{\Delta_{3/2}} =  m^{(1)}_{N_{5/2}}  = m^{(1)}_{\Delta_{5/2}} = m^{(1)}_{\Delta_{7/2}} ,
\end{equation}
\begin{equation}\label{mass4}
m_3 = m^{(2)}_{\Delta_{3/2}} =  m^{(2)}_{N_{5/2}} =  m^{(2)}_{\Delta_{5/2}} =  m^{(1)}_{N_{7/2}} =  m^{(2)}_{\Delta_{7/2}} = m^{(1)}_{\Delta_{9/2}},
\end{equation}
\begin{equation}\label{mass5}
m_4 = m^{(3)}_{\Delta_{5/2}} =  m^{(2)}_{N_{7/2}} = m^{(3)}_{\Delta_{7/2}} =  m^{(1)}_{N_{9/2}}  = m^{(2)}_{\Delta_{9/2}}
= m^{(1)}_{\Delta_{11/2}}.
\end{equation}

The masses (\ref{mass3})-(\ref{mass5}) are indicated in Column 2 of Tables \ref{amplitude2}-\ref{amplitude6}
for comparison with results obtained below from the resonance picture amplitudes. Here, the notation $m_K$ (or $m'_K$) is used 
for the calculated masses
while in Ref. \cite{COLEB2} the $m_K$ associated with $\ell =3$ are generic names  related to poles in the 
reduced amplitudes. One can notice that $m_2$ found in Ref.  \cite{COLEB1} for $\ell = 1$, as reproduced in 
Eq.  (\ref{clmass}),  is different from $m'_2$ obtained here for $\ell =3$.
In addition to distinct analytical forms for $m_2$ and $m'_2$ the 
coefficients $c_i$ entering these expressions are expected to 
depend on the band \cite{Matagne:2005gd}. A more extensive discussion 
is given at the end of the next section.

\section{The meson-nucleon scattering picture}

Here we are concerned with the SU(4) case, as above, and we look for the degeneracy pattern in the resonance picture.
Following Refs. \cite{COLEB1,COLEB2} the starting point in this analysis   are the linear relations 
of the S matrices  $S^{\pi}_{LL'RR'IJ}$ and  $S^{\eta}_{LRJ}$ of $\pi$ and $\eta$ scattering off 
a ground state baryon in terms of $K$-amplitudes. They are given by the following equations 
\begin{equation}\label{pi}
S^{\pi}_{LL'RR'IJ} = \sum_K ( - 1)^{R'-R} \sqrt{(2R+1)(2R'+1)} (2K+1)
\left\{\begin{array}{ccc}
        K& I & J \\
	R' & L' & 1
      \end{array}\right\} 
 \left\{\begin{array}{ccc}
        K& I & J \\
	R & L & 1
      \end{array}\right\}  
      s^{\pi}_{KLL'}    
\end{equation}
and
\begin{equation}\label{eta}
S^{\eta}_{LRJ} = \sum_K \delta_{KL}\delta(LRJ) s^{\eta}_{K}
\end{equation}
in terms of the reduced amplitudes $s^{\pi}_{KL'L}$ and $s^{\eta}_{K}$ respectively. 

The notation is as follows. For $\pi$ scattering $R$ and $R'$ are the spin of the incoming and outgoing baryons 
respectively ($R$ =1/2 for $N$ and $R$ = 3/2 for $\Delta$), $L$ and $L'$ are the partial wave angular momentum of the
incident and final $\pi$ respectively (the orbital angular momentum $L$ of $\eta$ remains unchanged), 
$I$ and $J$ represent the total isospin and total angular momentum
associated with a given resonance (see column 1 of Tables \ref{amplitude2}-\ref{amplitude6}) and $K$ is the 
magnitude  of ${\it grand}$ ${\it spin}$ $\vec{K} = \vec{I} + \vec{J}$.
The $6j$ coefficients imply four triangle rules $\delta(LRJ)$, $\delta(R1I)$, $\delta(L1K)$ and 
$\delta(IJK)$.

These equations were first derived in the context  of the chiral soliton model 
\cite{HAYASHI,MAPE,MATTIS,MattisMukerjee}
where 
the mean-field breaks the rotational and isospin symmetries, so that $J$ and $I$ are not
conserved but the ${\it grand}$ ${\it spin}$  $K$ is conserved and excitations can be labeled by $K$.
These relations are exact in large $N_c$ QCD and are independent of any model assumption.

The meaning of Eq. (\ref{pi}) is that there are more amplitudes $S^{\pi}_{LL'RR'IJ}$ than there are $s^{\pi}_{KLL'}$
amplitudes.  The reason is that the $I J$ as well as the $R R'$ dependence is contained only  in the geometrical
factor containing the two $6j$ coefficients.  
Then, for example, in the $\pi N$ scattering, in order for a resonance to occur in one channel there 
must be a resonance in at least
one of the contributing amplitudes $s^{\pi}_{KLL'}$. But as $s^{\pi}_{KLL'}$ contributes
in more than one channel,  all these channels resonate at the same energy and this implies degeneracy
in the excited spectrum.  From the chiral soliton model there is no reason to suspect degeneracy 
between different $K$ sectors.

From the meson-baryon scattering relations  (\ref{pi}) and   (\ref{eta})
the following degenerate negative parity multiplets have been found for $\ell$ = 1 orbital 
excitations \cite{COLEB1}
\begin{equation}\label{K0}
N_{1/2}, ~ \Delta_{1/2}, ~~~ (s^{\eta}_0)
\end{equation} 
\begin{equation}\label{K1}
N_{1/2},~ \Delta_{1/2},~N_{3/2},~ \Delta_{3/2},~ \Delta_{5/2}, ~~~ (s^{\pi}_{100}, s^{\pi}_{122})
\end{equation} 
\begin{equation}\label{K2}
\Delta_{1/2},  ~ N_{3/2},   ~ \Delta_{3/2}, ~ N_{5/2},  ~ \Delta_{5/2},   ~ \Delta_{7/2},  ~~~  (s^{\pi}_{222}, s^{\eta}_2).
\end{equation} 
One can see a clear correspondence
between the first three degenerate multiplets of Eqs. (\ref{K0}), (\ref{K1}) and (\ref{K2})
and  the three towers of states \cite{Pirjol:2003ye,COLEB1}
 of the excited quark picture provided by the 
symmetric core + excited quark scheme \cite{CCGL}.
They correspond to $K = 0, 1$ and 2 in the resonance picture.
But the resonance picture also provides a $K = 3$ due to the amplitude 
$s^{\pi}_{322}$. 
As this is different from the other $s^{\pi}_{KL'L}$ , in Ref. \cite{COLEB1}
it was interpreted as belonging to the $N = 3$ band. 

\begin{table}
\caption{
Partial wave  amplitudes and their expansions in
terms of $K$-amplitudes from   Eqs.~(\ref{pi}) and (\ref{eta}).
The superscripts $\pi N N$, $\pi N \Delta$, $\pi \Delta \Delta$, $\eta
N N$, and $\eta \Delta \Delta$ refer to the scattered meson and the
initial and final baryons, respectively.  
We list amplitudes consistent with a single quark
excited to $\ell=3$ and partial waves having $L = L^\prime  = 2 $.}
\renewcommand{\arraystretch}{1.25}
\label{amplitude2}
\medskip
\begin{tabular}{lcccccl}
\hline\hline
State \mbox{  } && Quark-shell mass \mbox{   } &&
\multicolumn{3}{l}{Partial wave and $K$-amplitudes} \\
\hline
$\Delta_{1/2}$ & & $m'_2$ & & $D_{31}^{\pi \Delta \Delta}$ &=&  $\frac{1}{10}\left(s^\pi_{122}+9 s^\pi_{222}\right)$\\
               & &        & & $D_{31}^{\eta \Delta \Delta}$  &=&  $s^\eta_2$ \\  
$N_{3/2}$ 
& & $m'_2$ & & $D_{13}^{\pi N N}$  &=& $ \frac{1}{2}
\left( s^\pi_{122} + s^\pi_{222}\right) $ \\
&& &&    $D_{13}^{\eta N N}$ & = & $s^\eta_2$ \\ 
& &  & &  $D_{13}^{\pi \Delta \Delta}$  &=& $ \frac{1}{2}
\left( s^\pi_{122} + s^\pi_{222}\right) $ \\
& &  & &  $D_{13}^{\pi N \Delta}$  &=& $ \frac{1}{2}
\left( s^\pi_{122} - s^\pi_{222}\right) $ \\
%
$\Delta_{3/2}$ 
&&  $m'_2, ~m_3$  && $D_{33}^{\pi N N}$  &=& $\frac{1}{20}
\left( s^\pi_{122} + 5 s^\pi_{222} + 14  s^\pi_{322} \right)$ \\
&&    && $D_{33}^{\pi \Delta \Delta}$  &=& $\frac{1}{25}
\left( 8 s^\pi_{122} + 10 s^\pi_{222} + 7  s^\pi_{322} \right)$ \\
&& &&    $D_{33}^{\eta \Delta \Delta}$ & = & $s^\eta_2$ \\ 
&&    && $D_{33}^{\pi N \Delta}$  &=& $\frac{1}{5 \sqrt{10}}
\left( 2 s^\pi_{122} + 5 s^\pi_{222} - 7  s^\pi_{322} \right)$ \\
%
$N_{5/2}$
&&  $m'_2, ~m_3$  && $D_{15}^{\pi N N}$  &=& $\frac{1}{9}
\left( 2 s^\pi_{222} + 7 s^\pi_{322}  \right)$ \\
&& && $D_{15}^{\eta N N}$ &=& $s^\eta_2$ \\
&&   && $D_{15}^{\pi \Delta \Delta}$  &=& $\frac{1}{9}
\left( 7 s^\pi_{222} + 2 s^\pi_{322}  \right)$ \\
&&    && $D_{15}^{\pi N \Delta}$  &=& $\frac{\sqrt{14}}{9}
\left( s^\pi_{222} -   s^\pi_{322} \right)$ \\
%
$\Delta_{5/2}$
&&  $m'_2, ~m_3$  && $D_{35}^{\pi N N}$  &=& $\frac{1}{90}
\left( 27 s^\pi_{122} +  35 s^\pi_{222} + 28 s^\pi_{322}  \right)$ \\
&&   && $D_{35}^{\pi \Delta \Delta}$  &=& $\frac{1}{450}
\left( 189 s^\pi_{122} + 5 s^\pi_{222} + 256 s^\pi_{322}  \right)$ \\
&& && $D_{35}^{\eta \Delta \Delta}$ &=& $s^\eta_2$ \\
&&    && $D_{35}^{\pi N \Delta}$  &=& $\frac{1}{90}\sqrt{\frac{7}{5}}
\left( 27 s^\pi_{122} + 5 s^\pi_{222} -  32 s^\pi_{322} \right)$ \\
%
$\Delta_{7/2}$
&&  $m'_2, ~m_3$  &&  $D_{37}^{\pi \Delta \Delta}$  &=& $\frac{1}{5}
\left( 2 s^\pi_{222} + 3 s^\pi_{322}  \right)$ \\
&& && $D_{37}^{\eta \Delta \Delta}$ &=& $s^\eta_2$ \\
[0.5ex]
  \hline\hline
\end{tabular}
\end{table}

\begin{table}\label{amplitude4}
\caption{
Same as Table \ref{amplitude2} but for partial waves  $L = L^\prime  = 4 $. }
\medskip \renewcommand{\arraystretch}{1.25}
\begin{tabular}{lcccccl}
\hline\hline
State \mbox{  } && Quark-shell mass \mbox{   } &&
\multicolumn{3}{l}{Partial wave and $K$-amplitudes} \\
\hline
$N_{5/2}$ 
& & $m_3$ & &  $G_{15}^{\pi \Delta \Delta}$  &=& $ s^\pi_{344} 
$ \\
%
$\Delta_{5/2}$ 
&&  $m_3,~m_4$  && $G_{35}^{\pi \Delta \Delta}$  &=& $\frac{1}{4}
\left( s^\pi_{344} + 3 s^\pi_{444}  \right)$ \\
&& && $G_{35}^{\eta \Delta \Delta}$ &=& $s^\eta_4$ \\
%
$N_{7/2}$ 
&&   $m_3,~m_4$ && $G^{\pi N N}_{17}$            &=& $\frac{1}{12}
\left( 7 s^\pi_{344} + 5 s^\pi_{444}  \right)$ \\ 
&& && $G_{17}^{\eta N N}$ &=& $s^\eta_4$ \\
&& && $G_{17}^{\pi \Delta \Delta}$  &=& $\frac{1}{12}
\left( 5 s^\pi_{344} + 7 s^\pi_{444}  \right)$ \\ 
&& && $G_{17}^{\pi N \Delta}$  &=& $\frac{\sqrt{35}}{12}
\left( s^\pi_{344} - s^\pi_{444}  \right)$ \\ 
$\Delta_{7/2}$

&&   $m_3,~m_4$    && $G^{\pi N N}_{37}$            &=& $\frac{1}{72}
\left( 7 s^\pi_{344} + 21 s^\pi_{444} + 44 s^\pi_{544} \right)$ \\
&& && $G_{37}^{\pi \Delta \Delta}$ &=& $\frac{1}{225}
\left( 100 s^\pi_{344} + 48 s^\pi_{444} + 77 s^\pi_{544} \right)$ \\
&& &&    $G_{37}^{\eta \Delta \Delta}$ & = & $s^\eta_4$ \\
&& && $G_{37}^{\pi N \Delta}$  &=& $\frac{\sqrt{14}}{90}
\left( 5 s^\pi_{344} + 6 s^\pi_{444} - 11 s^\pi_{544} \right)$ \\ 
$N_{9/2}$ 
&& $m_4$ && $G^{\pi N N}_{19}$            &=& $\frac{1}{15}
\left( 4 s^\pi_{444} + 11 s^\pi_{544}  \right)$ \\ 
&& && $G_{19}^{\eta N N}$ &=& $s^\eta_4$ \\
 & & & & $G_{19}^{\pi \Delta \Delta}$  &=& $ \frac{1}{15}
\left( 11  s^\pi_{444} + 4 s^\pi_{544}  \right)$ \\
&& && $G_{19}^{\pi N \Delta}$  &=& $\frac{ 2 \sqrt{11}}{15}
\left(  s^\pi_{444}  -  s^\pi_{544} \right)$ \\ 
$\Delta_{9/2}$ 
&&   $m_3,~m_4$ && $G^{\pi N N}_{39}$            &=& $\frac{1}{90}
\left( 35  s^\pi_{344} + 33 s^\pi_{444} + 22 s^\pi_{544}  \right)$ \\ 
& & & & $G_{39}^{\pi \Delta \Delta}$  &=& $ \frac{1}{900}
\left( 385  s^\pi_{344} +  3 s^\pi_{444} +  512 s^\pi_{544}  \right)$ \\
&& && $G_{39}^{\eta \Delta \Delta}$ &=& $s^\eta_4$ \\
&& && $G_{39}^{\pi N \Delta}$  &=& $\frac{1}{90} \sqrt{\frac{11}{10}}
\left(  35 s^\pi_{344} - 3 s^\pi_{444}  -  32 s^\pi_{544} \right)$ \\ [0.5ex]
$\Delta_{11/2}$ 
 & &$m_4$ & & $G_{3,11}^{\pi \Delta \Delta}$  &=& $\frac{1}{25}
\left(12 s^\pi_{444} + 13 s^\pi_{544} \right)$ \\
 & & & & $G_{3,11}^{\eta \Delta \Delta}$  &=&$s^\eta_4$ \\ 
  \hline\hline
\end{tabular}
\end{table}


\begin{table}\label{amplitude6}
\caption{
Same as Table \ref{amplitude2} but for partial waves  $L = L^\prime  = 6 $. }
\medskip \renewcommand{\arraystretch}{1.25}
\begin{tabular}{lcccccl}
\hline\hline
State \mbox{  } && Quark-shell mass \mbox{   } &&
\multicolumn{3}{l}{Partial wave and $K$-amplitudes} \\
\hline
$N_{9/2}$ 
 & & $m_4$ & & $I_{19}^{\pi \Delta \Delta}$  &=& $ s^\pi_{566}  $ \\

$\Delta_{9/2}$ 
& &  $m_3, ~m_4$  & & $I_{39}^{\pi \Delta \Delta}$  &=& $ \frac{1}{10}
\left( 3  s^\pi_{566} +  7 s^\pi_{666}  \right)$ \\
&& && $I_{39}^{\eta \Delta \Delta}$ &=& $s^\eta_6$ \\
$\Delta_{11/2}$ 
 & & $m_4$ & & $I_{1,11}^{\pi N N}$  &=& $\frac{1}{468}
\left( 55 s^\pi_{566} + 143 s^\pi_{666} + 270 s^\pi_{766} \right) $\\
 & & & & $I_{1,11}^{\pi \Delta \Delta}$  &=& $\frac{1}{819}
\left( 392 s^\pi_{566} + 130 s^\pi_{666} +297 s^\pi_{766} \right) $\\
&& && $I_{1,11}^{\eta \Delta \Delta}$ &=& $s^\eta_6$ \\
 & & & & $I_{1,11}^{\pi  N \Delta}$  &=& $\frac{\sqrt{55}}{117 \sqrt{14}}
\left( 14 s^\pi_{566} + 13 s^\pi_{666} - 27 s^\pi_{766} \right) $\\[1.1ex]
  \hline\hline
\end{tabular}
\end{table}

Here we extend the work of Ref. \cite{Pirjol:2003ye,COLEB1} to $\ell = 3$ excited states which 
belong to the $N = 3$ band. 
In Tables \ref{amplitude2}-\ref{amplitude6} we list the partial wave 
amplitudes of interest and their expansion  
in terms of $K$-amplitudes from Eqs.~(\ref{pi}) and (\ref{eta}). They correspond   
to   $L = L' = 2$,   $L = L' = 4$ and $L = L' = 6$ respectively. 
Note that the squared sum of the coefficients of  every elastic amplitudes    $\pi NN$ or $\pi \Delta \Delta$
is equal to one.
This is due to  the sum rule of $6j$ coefficients
\begin{equation}\label{sumrule}
\sum_K  (2R+1) (2K+1)
\left\{\begin{array}{ccc}
        K& I & J \\
	R' & L & 1
      \end{array}\right\} 
 \left\{\begin{array}{ccc}
        K& I & J \\
	R & L & 1
      \end{array}\right\}    
      = \delta(R'R),
\end{equation}
which can be used for a check. The same relation can be used to check that the coefficients of 
the  $\pi N \Delta$ amplitudes sum up to zero.

From the last column of Tables  \ref{amplitude2}-\ref{amplitude6}  one can infer the following degenerate towers of states
with their contributing amplitudes 
\begin{eqnarray}
\Delta_{1/2}, \; \; \;   N_{3/2} , \; \; \;   \Delta_{3/2} , \; \; \;   N_{5/2} , \; \; \;  \Delta_{5/2} , \; \; \;  \Delta_{7/2} , \; \; \;
&~&
 (s_{222}^\pi, s_{2}^\eta) , \label{s2p}\\
 \Delta_{3/2} , \; \; \;   N_{5/2} , \;  \; \;  \Delta_{5/2} , \; \;  \; N_{7/2} , \; \; \;
\Delta_{7/2} , \; \; \;  \Delta_{9/2} , 
 &~& (s_{3 2 2}^\pi,  s_{3 4 4}^\pi) , \label{s1}\\
 \Delta_{5/2} , \;  \; \;  N_{7/2} , \;  \; \; \Delta_{7/2} , \; \; \; N_{9/2} , \; \; \;  \Delta_{9/2} ,
 \;  \; \; \Delta_{11/2} , 
 &~& ( s_{4 4 4}^\pi, s_{4}^\eta ) , \label{s2} \\
\Delta_{7/2} , \; \; \; N_{9/2} , \; \;  \; \Delta_{9/2} ,
 \; \; \; \Delta_{11/2} , \; \; 
&~&
(s^\pi_{5 4 4 },  s^\pi_{5 6 6 }), \;  \;  \label{s3} \\
\Delta_{9/2}, \; \; \; 
\Delta_{11/2} , 
&~& (s^\pi_{6 6 6 }, s_{6}^\eta ) \label{s4}
\end{eqnarray}
associated with  $K = 2, 3, 4, 5$ and 6 respectively.
Here one can recognize  patterns of degeneracy similar to those observed in  Table II of Ref. \cite{COLEB2}. 
Note  that $m_K$ of column 2 of that table represents the name associated with the position of a possible pole in 
an amplitude with $K$-spin. 

We can now  compare the towers (\ref{s2p})-(\ref{s4})
 with the quark-shell model results of (\ref{mass3})-(\ref{mass5}).
The first observation is that  the agreement of 
 (\ref{s2p}) ($K = 2$) with (\ref{mass3}), of
 (\ref{s1}) ($K = 3$) with  (\ref{mass4})  and  of (\ref{s2})  ($K = 4$)  with (\ref{mass5}) 
is perfect regarding the quantum numbers.
Second, we note that the resonance picture can have poles with $K = 5, 6$ 
 which imply the towers (\ref{s3}) and (\ref{s4}).  They have no counterpart
in the quark-shell picture for $\ell = 3$. 
But there is no problem because the poles with $K = 5, 6$  can belong to a higher band, 
namely $N = 5$ ($\ell = 5$) without spoiling the compatibility. 

A discussion is necessary for the  tower (\ref{mass3}) of the quark-shell picture  associated with the
degenerate mass $m'_2$. The expression of  $m'_2$ is entirely different from that of $m_2$
obtained for $\ell$ = 1. This is  quite natural from the algebra described in the Appendix.
Moreover, in practice, the $\ell$ = 3 states should lie higher
than the $\ell$ = 1 states (as they include more orbital excitation). 
In fact the analysis of Ref.  \cite{Matagne:2005gd} suggests that the coefficients $c_i$ 
are expected to depend  on the band.
If so, some constraints could be imposed on the values of these  coefficients  to be 
found phenomenologically, after including $1/N_c$ corrections. 
For example, the first panel in Fig. 1 of Ref.  \cite{Matagne:2005gd} indicates 
a linear behavior of the coefficient $c_1$  as a function of the band N. From that figure one
can extract the value of $c_1$ associated with the band N = 3. This gives
\begin{equation}\label{paramc1}
c_1 \approx 640~ {\rm MeV}
\end{equation}
Such a value can safely be used in a first step analysis of the N = 3 band, which would mean that 
there is one less parameter to fit. 

Thus one can associate a common $K = 2$ to $\ell = 1$ and $\ell = 3$.  For this value of $K$ 
the triangular rule $\delta( K \ell 1)$ proposed in Ref \cite{COLEB2} is satisfied.
The quark-shell picture brings however more information than the resonance picture
because it implies  an energy dependence via the $\ell$ dependence which
measures the orbital excitation.  As $m'_2$ is different from $m_2$ and in the resonance 
picture they stem from the same amplitude $s^{\pi}_{222}$ one should expect that this
amplitude possesses two poles at two distinct energies, in order to have compatibility.
Thus the number of poles of the reduced amplitudes $s^{\pi}_{KLL}$ remains an open question.

We anticipate that a similar situation will appear  for every value of $K$ 
associated with two distinct values of $\ell$, satisfying the $\delta(K \ell 1)$ rule, for example, for
$K$ = 4 which is  common to $\ell$ = 3  and $\ell$ = 5.

\section{The experimental situation for resonances with $\ell \geq$ 3}

\begin{table}
\caption{Examples of nonstrange negative parity resonances from Particle Data Group 
 \cite{PDG}  and their possible main component  expressed in terms of SU(6) multiplets.}
\renewcommand{\arraystretch}{1.5}
\begin{tabular}{crr}
\hline
\hline
           Resonance  & \ \ \ \ \ \ \ Multiplet  & \ \ \ \ \ \ \ Status  \\
\hline
$N(2190) G_{17}$ & $^2 N [70, 3^-]$ & $****$  \\
$N(2250) G_{19}$ & $^4 N [70, 3^-]$ &$ ****$    \\
$N(2600)  I_{1,11}$ &   $^2 N [70, 5^-]$ &  $***$  \\
$\Delta(2220) G_{37}$ & $ ^2 N [70, 3^-]$ & $*$  \\
$\Delta(2400) G_{39}$ &  $^4 \Delta [56, 3^-]$ & $**$  \\
$\Delta(2750) I_{3,13}$ &  $^4 \Delta [56, 5^-]$ & $**$  \\[0.5ex]
\hline
\hline
\end{tabular}

\label{exp}
\end{table}

Here we are essentially concerned with  resonances which can be explained as orbital 
excitations  with $\ell$ = 3. 
Examples of experimentally known negative parity resonances of this 
category \cite{PDG}  are indicated in Table \ref{exp}. 
They are located at about 2.2 GeV and in quark model terms they belong to the N = 3 band. 
For completeness we have also indicated two resonances $N(2600)  I_{1,11}$ and $\Delta(2750) I_{3,13}$ 
which should belong to the N = 5 band, as their total angular momentum require an orbital excitation with $\ell$ = 5.

The  SU(6) $\times$ O(3) multiplet  content  of the N = 3 band is \cite{Dalitz1977,Stancu:1991cz} :
$[70', 1^{-}], [70'', 1^{-}], [56, 3^{-}], [20, 3^{-}], [70, 3^{-}],
[70, 2^{-}], [56, 1^{-}]$ and $[20, 1^{-}]$ where  $70'$ and $70''$ represent radial
excitations.  In column 2  we have indicated the multiplet to which 
the listed resonances can belong,  using the notations of Ref. \cite{Stancu:1991cz}.
Both references \cite{Dalitz1977} and \cite{Stancu:1991cz} were independently concerned 
with  the resonance $D_{35}$ observed by Cutkosky et al. \cite{Cutkosky}.
In Ref. \cite{Dalitz1977} a sum rule was derived in a harmonic oscillator basis 
to calculate the mass of  $D_{35}$ as a pure $[56, 1^{-}]$ state by neglecting the tensor force. 
 In  Ref. \cite{Stancu:1991cz}  the masses of  negative parity nonstrange resonances were obtained in a quark model with 
a linear confinement and a chromomagnetic interaction
(spin-spin and tensor forces) including  interband mixing.  The $D_{35}$ resonance was found to be mainly 
a $[56, 1^{-}]$ state.
 
For the resonances expected to belong to
the N = 5 band the multiplet is only suggested. To our knowledge, the SU(6) $\times$ O(3)  multiplet decomposition 
of this band is  not known.

\section{Conclusions}

The compatibility between the quark-excitation and the meson-nucleon resonance
pictures of negative parity baryons with $\ell$ = 3 has been analyzed in the spirit of 
Refs. \cite{Pirjol:2003ye,COLEB1,COLEB2}. 
We have found patterns of degeneracy with a common resonance content in both pictures.
This supports the idea of full compatibility of Ref. \cite{COLEB2} in the sense that any complete spin-flavor multiplet within one picture fills the quantum numbers of the
other picture. However the quark-shell picture is richer in information, by making a clear distinction 
between degenerate sets of states of different values of the angular momentum but
associated with the same grand spin $K$.
  
The low-energy baryons 
of the $[70,1^-]$ multiplet have been very extensively studied in large $N_c$ QCD but the highly excited
$\ell $ = 3 baryons have been nearly entirely neglected so far. To our knowledge, there is  
only one  general work including   $\ell $ = 3 baryons  \cite{Goity:2007sc}.
 The experimental situation described 
in Sec. 4 encourages such analysis.  The $1/N_c$ expansion method could, in principle,  
predict many more resonances to guide 
the experimentalists. 
This work supplies an incentive for the study of highly excited negative parity baryon 
in the $1/N_c$ expansion method. Including $1/N_c$ corrections in the mass formula means
that, besides $c_1$, $c_2$  and $c_3$,  more parameters are involved in the fit. 
As the data is presently scarce, in a first attempt, the number of parameters must remain small.
A strategy would be to fix the value of $c_1$ in agreement  with
Eq. (\ref{paramc1}) and restrict the number of operators 
in the mass formula to the most dominant 
ones, as for example,  the spin and isospin 
operators  described in Ref.  \cite{Matagne:2008fw}  for $\ell$ = 1.   
This would involve at most four parameters to fit. 
\appendix
\section{}
In our notation \cite{Matagne:2008fw} the total wave function of the symmetric core+excited quark procedure \cite{CCGL}
takes  the following form 
\begin{equation}\label{wfp}
|\ell S J J_3; I I_3 \rangle_{p=2}  = 
\sum_{m_\ell,S_3} 
   \left(\begin{array}{cc|c}
	\ell    &    S   & J   \\
	m_\ell  &    S_3  & J_3 
      \end{array}\right)
|\ell m \rangle       
|[N_c-1,1]p=2;S S_3; I I_3 \rangle.
\end{equation}
It contains a Clebsch-Gordan coefficient, an orbital part $|\ell m \rangle$  and  a  spin-flavor part $|[N_c-1,1]p=2;S S_3; I I_3 \rangle$
depending on an index $p$ which takes the value 2, which signifies that the 
excited quark is in the second row of the Young diagram of partition $[N_c-1,1]$ in the 
flavor-spin space. The expression of the spin-flavor part  is 
\begin{eqnarray}\label{fs}
|[N_c-1,1]p =2;S S_3; I I_3 \rangle = 
\sum_{p_1 p_2} K([f_1]p_1[f_2]p_2|[N_c-1,1]p=2) 
|S S_3;p_1 \rangle |I I_3;p_2 \rangle,   
\end{eqnarray}
in terms of the spin part 
\begin{equation}\label{spin}
|S S_3; p_1 \rangle = \sum_{m_1,m_2}
 \left(\begin{array}{cc|c}
	S_c    &    \frac{1}{2}   & S   \\
	m_1  &         m_2        & S_3
      \end{array}\right)
      |S_cm_1 \rangle |1/2m_2 \rangle,
\end{equation}
with $S_c = S - 1/2$ if $p_1 = 1$    and $S_c = S + 1/2$ if   $p_1 = 2$ and the isospin part 
\begin{equation}\label{isospin}
|I I_3; p_2 \rangle = \sum_{i_1,i_2}
 \left(\begin{array}{cc|c}
	I_c    &    \frac{1}{2}   & I   \\
	i_1    &       i_2        & I_3
      \end{array}\right)
      |I_c i_1 \rangle |1/2 i_2 \rangle,
\end{equation}
with $I_c = I - 1/2$ if $p_2 = 1$   and   $I_c = I + 1/2$  if $p_2 = 2$. 
Here    $S_c$ and $I_c$ are the spin and isospin of the core and  $ p_1 $ 
and $p_2$ represent the position of the $N_c$-th quark in the spin   and 
isospin   parts of the wave function respectively, both consistent 
with $p$ = 2  and the inner product rules generating the wave function in the 
flavor-spin space.   
The coefficients $K([f_1]p_1[f_2]p_2|[N_c-1,1]p=2)$ are isoscalar factors of the permutation group $S_{N_c}$.
At $p$  fixed, one can use an alternative notation  
\begin{equation}
K([f_1]p_1[f_2]p_2|[N_c-1,1]p=2) = c^{[N_c-1,1]}_{p_1 p_2} (S)
\end{equation}

For the representation $[N_c-1,1]$ the only non vanishing expressions are 
\begin{equation}\label{c11}
c^{[N_c-1,1]}_{1 1} (S) = - \sqrt{\frac{(S+1)(N_c - 2S)}{N_c (2S+1)}},
\end{equation}
\begin{equation}\label{c22}
c^{[N_c-1,1]}_{2 2} (S) = \sqrt{\frac{S[(N_c + 2(S+1)]}{N_c (2S+1)}},
\end{equation}
\begin{equation}\label{c12}
c^{[N_c-1,1]}_{1 2} (S) = c^{[N_c-1,1]}_{2 1} (S) =1 .
\end{equation}

Actually we need the above coefficients in the limit $N_c \rightarrow \infty$. Therefore for $N$ resonances
where $S$ = 1/2 we have to take 
\begin{equation}\label{Ninfty}
c^{[N_c-1,1]}_{1 1} (1/2) \rightarrow  - \sqrt{\frac{3}{4}} ; ~~~~~~c^{[N_c-1,1]}_{2 2} (1/2) \rightarrow  \sqrt{\frac{1}{4}},
\end{equation}
and for $\Delta$ resonances 
\begin{equation}\label{Dinfty}
c^{[N_c-1,1]}_{1 1} (3/2) \rightarrow  - \sqrt{\frac{5}{8}} ; ~~~~~~c^{[N_c-1,1]}_{2 2} (3/2) \rightarrow  \sqrt{\frac{3}{8}}.
\end{equation}

With the above notations, the matrix elements with a given $J$, between states with $S'$ and $S$ take the following form
\begin{eqnarray}
\lefteqn{\langle \ell' S'J'J'_3;I'I'_3 | \ell \cdot s | \ell S JJ_3; I I_3 \rangle_{p=2}  = }\nonumber \\& & ( -1)^{J+\ell+1/2} \delta_{J'J} \delta_{J'_3J_3}\delta_{\ell'\ell} \delta_{I'I} \delta_{I'_3I_3} \sqrt{\frac{3}{2}(2 S + 1) (2 S' + 1) \ell (\ell + 1)(2 \ell + 1)} 
 \left\{\begin{array}{ccc}
        \ell & \ell & 1 \\
	S & S' & J
      \end{array}\right\} 
\nonumber \\& &\times \sum_{p_1,p_2} ( -1)^{-S_c} c^{[N_c-1,1]}_{p_1p_2}(S') c^{[N_c-1,1]}_{p_1p_2}(S)
 \left\{\begin{array}{ccc}
        1 & \frac{1}{2} &  \frac{1}{2}\\
	S_c & S & S'
      \end{array}\right\}  
\end{eqnarray}
This expression is equivalent to  Eq. (A7) of Ref. \cite{CCGL}.  The correspondence in the isoscalar factors denoted there by $c_{\rho \eta}$ is
\begin{equation}\label{coresp}
       c^{[N_c-1,1]}_{1 1} (S) \rightarrow c_{0-}; ~~~~c^{[N_c-1,1]}_{2 2} (S) \rightarrow c_{0+};
~~~ c^{[N_c-1,1]}_{1 2} (S) \rightarrow c_{++}; ~~~~c^{[N_c-1,1]}_{2 1} (S) \rightarrow c_{--}.
\end{equation}
The expectation value of the  operator  containing the tensor term  is
\begin{eqnarray}
\lefteqn{\langle\ell' S'J'J'_3;I'I'_3 | \ell^{(2)} \cdot g \cdot G_c  | \ell S JJ_3; I I_3 \rangle_{p=2}  =(-)^{J+I+\ell+S+1/2}} \nonumber \\& &\times  \delta_{J'J} \delta_{J'_3J_3}\delta_{\ell'\ell} \delta_{I'I} \delta_{I'_3I_3}
 \frac{1}{8} \sqrt{\frac{15}{2} \ell (\ell+1)(2\ell-1)(2\ell+1)(2\ell+3) (2S'+1)(2S+1)  } 
 \left\{\begin{array}{ccc}
       2 & \ell & \ell  \\
	J &  S & S' 
      \end{array}\right\} \nonumber \\
& & \times   
\sum_{p'_1,p'_2,p_1,p_2}  ( -1)^{S'_c}  c^{[N_c-1,1]}_{p'_1 p'_2}(S') c^{[N_c-1,1]}_{p_1 p_2}(S)
\sqrt{(2 I'_c+1)(2 I_c+1)} \nonumber \\
& & \times  
\sqrt{(N_c+1)^2 - (S'_c - S_c)^2 (2 I + 1)^2}
 \left\{\begin{array}{ccc}
       \frac{1}{2} & 1 &  \frac{1}{2} \\
	I'_c          &  I & I_c 
      \end{array}\right\} 
 \left\{\begin{array}{ccc}
       I'_c            &  I_c & 1 \\
	S'             &  S   &  2 \\
       \frac{1}{2}  &  \frac{1}{2} & 1
      \end{array}\right\}       
\end{eqnarray} 
One can recover Eq. (A9) of Ref. \cite{CCGL} using  the correspondence (\ref{coresp}). 
In the large  $N_c$ limit considered here the term  $(S'_c - S_c)^2 (2 I + 1)^2$ 
under the squared root should be ignored.

\vspace{2cm} 
 
{\bf Acknowledgments}
 The work of one of us (N. M.) was supported by F.R.S.-FNRS (Belgium).



\end{document}